\documentclass[11pt]{article}
\usepackage{epsfig,dsfont}
\usepackage{geometry}
\begin{document}
\sffamily

\vspace*{1mm}

\begin{center}

{\LARGE
Canonical simulations with worldlines: \\
\vskip3mm
an exploratory study in $\phi^{4}_{2}$ lattice field theory}
\vskip15mm
Oliver Orasch and Christof Gattringer
\vskip5mm
Universit\"at Graz, Institut f\"ur Physik, Universit\"atsplatz 5, 8010 Graz, Austria 
\end{center}
\vskip12mm

\begin{abstract}
In this letter we explore the perspectives for canonical simulations in the worldline formulation of a lattice field theory. Using the charged 
$\phi^4$ field in two dimensions as an example we present the details of the canonical formulation based on worldlines and outline algorithmic strategies for canonical worldline simulations. We discuss the steps for converting the data from the canonical approach 
to the grand canonical picture which we use for cross-checking our results. The canonical approach presented here can easily 
be generalized to other lattice field theories with a worldline representation.
\end{abstract}

\vskip14mm

\section{Introductory remarks}

Implementing a clean ab-initio calculation of lattice QCD at finite density is one of the great open challenges in the field. The reason 
for the difficulties is the fact that at non-zero chemical potential $\mu$ the action $S$ becomes complex and the Boltzmann factor 
$\exp(-S)$ cannot be used as a weight factor in a Monte Carlo simulation. Among the different approaches for overcoming that 
so-called complex action problem are canonical simulations where one works at fixed net baryon number, i.e., at a fixed density. 
The key challenge for implementing the canonical strategy in QCD is the projection to the desired quark or baryon number: 
in the conventional path integral representation the quark number is not an integer valued observable and the fermion
determinant has to be decomposed into temporal winding classes for the gauge loops it consists of. Various strategies for that task 
can be found in the literature. They either use different expansions of the fermion determinant or Fourier transformation of the determinant
with respect to imaginary chemical potential (see \cite{canonicalQCD} for some examples). 
However, so far the results for canonical lattice QCD are restricted to small 
volumes and low densities \cite{canonicalQCD}. 

In recent years an alternative formulation based on worldlines was found and explored for several lattice field theories 
(see \cite{reviews} for reviews given at the yearly lattice conference series on that topic). In the worldline (or dual) formulation
the partition function $Z$ is exactly rewritten in terms of new variables, such that $Z$ becomes a sum over configurations
of worldlines. The corresponding weights for the worldline configurations 
are real and positive also at non-zero chemical potential and the complex action 
problem is overcome completely in such a worldline representation. 

So far the worldline approach has only been used in the grand canonical formulation, but as a matter of fact, the worldline formulation 
is very well suited also for the canonical approach which we develop here. Due to certain topological constraints of the worldlines
(see the comments in our summary in Section 5) in some cases the canonical worldline approach is expected to  
outperform the grand canonical one. The canonical approach is very natural in the worldline formulation because 
the particle number is given by the winding number of the worldlines around the compactified time direction and thus always is an
integer. It is straightforward to construct algorithms for simulating the worldlines that do not change the winding number such 
that the system is treated canonically, i.e., at fixed particle number. 

In this letter we develop the canonical worldline approach for the simple example of the charged scalar $\phi^4$ field in two dimensions
($\phi^{4}_{2}$ lattice field theory), but already stress here that the generalization to other lattice field theories with a worldline
representation is straightforward. We discuss two different update strategies for the canonical simulation, as well as the steps for
converting the canonical results to the grand canonical picture. We use this construction to self-consistently cross-check 
canonical and grand canonical results for our observables in order to evaluate the new approach.

\section{The model and its worldline representation}

As already outlined in the introduction we develop the canonical worldline approach using the charged $\phi^4$ field 
in two dimensions as an example. The Euclidean action for $\phi^{4}_{2}$ lattice field theory in the conventional grand 
canonical formulation reads
\begin{equation}
S[\phi]  \; = \; 
\sum_{x \in \Lambda} \left[ (4 + m^{2}) \, |\phi _{x}|^2 + \lambda |\phi _{x}|^4 - \sum_{\nu = 1}^{2} 
\left( e^{\mu \delta_{\nu,2}}\phi _{x}^{\ast} \phi _{x+\hat{\nu}} + e^{-\mu \delta_{\nu,2}}\phi _{x}^{\ast} \phi _{x-\hat{\nu}} \right)\right] \; .
\label{action_conventional}
\end{equation}
The dynamical degrees of freedom are the complex valued fields $\phi_x \in \mathds{C}$ assigned to the sites $x$ of a
2-dimensional $N_s \times N_t$ lattice $\Lambda$ with periodic boundary 
conditions. The bare mass is denoted by $m$ and the coupling of the 
quartic self-interaction by $\lambda$. We introduce a chemical potential $\mu$, which in the lattice formulation 
gives a different weight for forward and backward propagation in the Euclidean time direction ($\nu = 2$).  The inverse temperature 
$\beta$ in lattice units is given by the temporal extent of the lattice, i.e., $\beta = N_t$. 
The grand canonical partition function of the system reads 
\begin{equation}
Z_{gc} \; = \; \int \! \mathcal{D}[\phi]  \; e^{-\, S[\phi]} \qquad \mbox{with} 
\qquad  \int \! \mathcal{D}[\phi] \; = \; \prod_{x \in \Lambda} \frac{1}{2\pi} \int_{\mathds{C}} \! d \phi_x \; .
\label{latticeZ}
\end{equation}
In the conventional representation (\ref{action_conventional}) the action $S[\phi]$ becomes complex for 
$\mu \neq 0$ and the Boltzmann factor $e^{-\, S[\phi]}$ in (\ref{latticeZ}) can not be used as a probability 
in a Monte Carlo simulation. In other words:
for finite chemical potential the model has a complex action problem in the conventional representation. 

However, the complex action problem of the model can be overcome by exactly mapping the partition function to a worldline 
representation where the partition sum has only real and positive contributions. The idea is to expand the Boltzmann factor
for the nearest neighbor term and to subsequently integrate out the original field variables $\phi_x$. When expanding the nearest 
neighbor Boltzmann factors integer valued summation variables that live on the links of the lattice need to be introduced, which 
often are referred to as "dual variables". Integrating out the original field variables gives rise to constraints for the dual variables
such that admissible configurations have the form of closed loops of worldlines. The integration over the $\phi_x$
generates real and positive weight factors that depend on the parameters $m, \lambda$ and $\mu$. We do not display the
derivation of the worldline representation here and only give the final result for the partition sum (for a derivation see, e.g., 
\cite{phi4}).

\begin{figure}[t]
\begin{center}
\includegraphics[height=5.3cm]{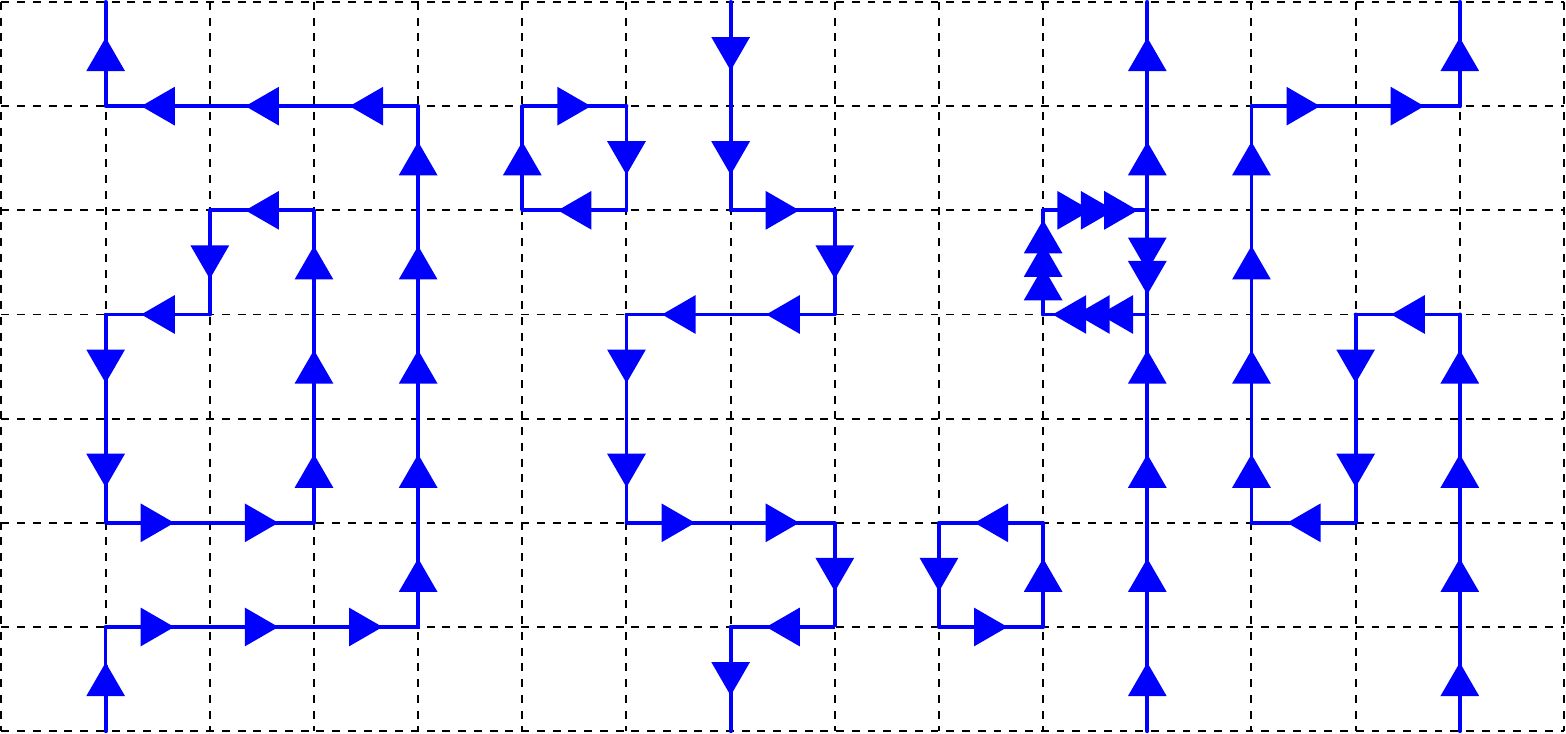}
\end{center}
\vspace*{-4mm}
\caption{Example of an admissible worldline configuration of the dual link variables $k_{x,\nu}$. The horizontal direction corresponds 
to space, while the vertical direction is Euclidean time. The configuration has a temporal net-winding number $W_t[k] = + 2$.}
\label{fluxlines}
\end{figure}

In the worldline representation the grand canonical partition function (\ref{latticeZ}) is exactly rewritten to
\begin{equation}
Z_{gc} \; = \; \sum_{\{k\}}  e^{\, \mu \beta \, W_t[k]} \; B[k] \; \prod_{x} \delta(\vec{\nabla} \cdot \vec{k}_{x}) \; .
\label{worldlineZ}
\end{equation}
Now the dynamical variables are the integer valued dual variables $k_{x,\nu} \in \mathds{Z}$ 
assigned to the links of the lattice, and by $\sum_{\{k\}}$ we denote the sum over all possible configurations of the $k_{x,\nu}$. 
As already remarked, the dual variables $k_{x,\nu}$ are subject to constraints which are represented by the product 
$\prod_{x \in \Lambda} \delta(\vec{\nabla} \cdot \vec{k}_{x})$ over all sites $x$ of the lattice, where $\delta(j) \equiv \delta_{j,0}$
is the Kronecker Delta. By $\vec{\nabla} \cdot \vec{k}_{x}$ we denote the lattice version of the divergence of $k_{x,\nu}$ defined as
\begin{equation}
\vec{\nabla} \cdot \vec{k}_{x} \; \equiv \;  \sum_{\nu}(k_{x,\nu} - k_{x-\hat{\nu},\nu}) \; .
\end{equation}
The constraint implies that the $k_{x,\nu}$ have vanishing divergence at all sites of the lattice. In other words: at every site $x$
the net flux of $k_{x,\nu}$ vanishes and the admissible configurations of the $k_{x,\nu}$ have the form of worldlines of conserved flux. 
An example of an admissible configuration is shown in Fig.~\ref{fluxlines} where a value of $k_{x,\nu} > 0$ ($< 0$) 
is represented by $\mid \! k_{x,\nu} \! \mid$ arrows on the link $(x,\nu)$ in positive (negative) $\nu$-direction. 

A particularly elegant aspect of the worldline representation is the fact that the net particle number $N$ is given by the net winding 
number $W_t[k]$ of the conserved $k$-flux around the compactified time direction. This property is immediately evident by comparing 
the standard expression $e^{ \, \mu \beta N}$ for the coupling of the net particle number $N$ with the term $e^{\, \mu \beta \, W_t[k]}$
that appears in (\ref{worldlineZ}), such that we read off the following worldline expression for the net particle number $N$ of a 
given configuration of $k$-worldlines: 
\begin{equation}
N \; = \; W_t[k] \; \in \mathds{Z} \; .
\label{NisW}
\end{equation}
The example shown in Fig.~\ref{fluxlines} is a worldline configuration with $W_t[k]  = +2$.

We stress at this point, that this beautiful geometrical 
integer valued expression for the net particle number is specific for the worldline representation. In the conventional representation 
the particle number is the spatial integral of the temporal component of the discretized vector current of the field $\phi_x$, and is 
not an integer for an arbitrary configuration of the $\phi_x$. The construction of the canonical approach described below crucially 
hinges on this integer valued net particle number $N = W_t[k]$.

To complete the discussion of the worldline form of the grand canonical partition sum (\ref{worldlineZ}) we still need to discuss
the real and positive weights factors $B[k]$. They are given by 
\begin{eqnarray}
&& \hspace{30mm}
B[k] \; = \; \sum_{\{a\}} \, \prod_{x,\nu}\frac{1}{(a_{x,\nu}+|k_{x,\nu}|)! \, a_{x,\nu}!} \,  \prod_{x} I(s_x) \qquad \mbox{with} 
\label{weights} \\
&& 
I(s_x) \; = \; \int_{0}^{\infty} d r \; r^{\, s_x + 1} \, e^{-(4 + m^{2}) \, r^2 - \lambda r^4} \quad \mbox{and} \quad
s_x \; = \; \sum_{\nu}\Big[|k_{x,\nu}| +  |k_{x-\hat{\nu}}| + 2(a_{x,\nu} +  a_{x-\hat{\nu}})\Big] \; .  \nonumber
\end{eqnarray}
The $B[k]$ are themselves given as a sum over configurations $\sum_{\{a\}}$ of integer valued auxiliary link variables 
$a_{x,\nu} \in \mathds{N}_0$. However, the $a_{x,\nu}$ are not subject to constraints and the sums for the
$B[k]$ can be treated with a conventional Monte Carlo simulation. The weight factors in (\ref{weights}) come from the  
expansion of the exponentials of the nearest neighbor term Boltzmann factors, as well as from integrating out the radial degrees 
of freedom of the original field variables. That latter contribution gives rise to the integrals $I(s_x)$, which in an actual 
numerical simulation are pre-calculated numerically and stored for the parameters $m$ and $\lambda$ one wants to simulate at and for a 
sufficient number of values of the integer valued arguments $s_x \geq 0$. These integers $s_x$ combine the absolute values
of the $k_{x,\nu}$ and $a_{x,\nu}$ variables at a given site$\;x$. In the worldline representation (\ref{worldlineZ}) -- (\ref{weights}) 
the partition sum has only real and positive contributions, such that the complex action problem is solved.

For the evaluation of observables the observables need to be expressed in terms of the worldlines. 
Here we are interested in the vacuum expectation value $\langle n \rangle$  of the net particle 
number density $n = N/N_s$ and the field expectation value $\langle |\phi |^2 \rangle$. When
comparing canonical and grand canonical results we will also need the quartic field expectation value 
$\langle |\phi |^4 \rangle$ for an intermediate step. These three observables can be 
obtained as derivatives of the free energy density $f$ which is defined as
\begin{equation}
f \; \equiv \; - \frac{1}{N_s \beta} \ln Z \; = \; - \frac{1}{N_s N_t} \ln Z\; .
\end{equation}
Here $Z$ can either be the grand canonical partition function $Z_{gc}$, or the canonical partition function $Z_N$ at fixed
net-particle number $N$ which we define below. The derivatives of $f$ with respect to the parameters can be applied also to the worldline
representation of $Z$, giving rise to the dual form of the observables (we here partly use $\beta = N_t$ and denote the lattice volume 
with $V = N_s \, N_t$),
\begin{equation}
\langle n \rangle  = - \frac{\partial f}{\partial \mu}  =    \frac{\langle W_t \rangle }{N_s} \; , \; \; 
\langle |\phi |^2 \rangle =  \frac{\partial f}{\partial m^2} =  \frac{1}{V} \!
\left\langle \! \sum_{x} \!\frac{I(s_x\!+\!2)}{I(s_x)} \! \right\rangle \; , \; \;
\langle |\phi |^4 \rangle = \frac{\partial  f}{\partial \lambda} =  \frac{1}{V} \!
\left\langle\! \sum_{x} \!\frac{I(s_x\!+\!4)}{I(s_x)} \! \right\rangle \, ,
\label{observables}
\end{equation}
where the vacuum expectation values on the right hand sides of the equations are now understood in the worldline representation. The dual
form of the observables are first moments of the winding number or of ratios of the weight factors $I(s_x)$, i.e, they are simple to evaluate 
for given worldline configurations.

In this paper we will compare the results from canonical and grand canonical simulations. Before we discuss the
canonical approach in the worldline representation, we briefly comment on the numerical simulation of the grand canonical ensemble
described by (\ref{worldlineZ}) -- (\ref{weights}). For a more detailed discussion see \cite{phi4worm}. In the 
worldline form we have two sets of dynamical variables living on the links: $k_{x,\nu} \in \mathds{Z}$ and
$a_{x,\nu} \in \mathds{N}_0$ (which can be viewed as generating the weight of the $k_{x,\nu}$ configurations). 
While the $k_{x,\nu}$ have to obey flux conservation, the $a_{x,\nu}$ are unconstrained and we update them with local
Metropolis sweeps. For the update of the $k_{x,\nu}$ we explored two strategies: local updates combined with updates of 
winding flux, as well as an update based on the Prokofev-Svistunov worm algorithm \cite{worm}. Both these update strategies can 
also be used for the canonical approach defined in the next section.

For the local updates we implemented sweeps where the $k$-flux around a plaquette is increased or decreased by one unit. 
This step is accepted with a Metropolis decision. In order to obtain an ergodic algorithm, we augmented the local updates with winding 
flux updates where we attempt to increase (decrease) by $+1$ ($-1$) 
the $k$-flux along straight loops that close around the compact space or 
time directions. Also these winding flux updates need to be accepted with Metropolis decisions, where for the case of
temporally winding loops the Metropolis probabilities depend also on the chemical potential (compare (\ref{worldlineZ})).
A combined sweep of the local/winding flux update consists of a sweep of local updates for all plaquettes together with a sweep
over all spatial and temporal straight loops. 

For the worm update one needs a slight generalization of the original worm algorithm \cite{worm}, since here we have 
weights on the links, but also on the sites. In \cite{phi4worm} it is shown how the site weights at the two endpoints of a
link visited by the worm can be distributed such that detailed balance is implemented correctly. Furthermore we used an additional
amplitude factor for the starting and terminating steps of the worm to optimize the performance in different regions of parameter
space (for details see \cite{phi4worm}).

\section{Setup of a canonical worldline calculation}

Having presented the conventional and the worldline representation of our model in the grand canonical picture, let us now come
to discussing the corresponding canonical form. The key ingredient for the canonical worldline formulation is
Eq.~(\ref{NisW}) which identifies the integer valued temporal winding number $W_t[k]$ of $k$-flux as the particle number $N$. Thus 
from (\ref{worldlineZ}) we can read off the following expression for the canonical partition function $Z_N$ with fixed net particle
number $N$:
\begin{equation}
Z_N \; = \; \sum_{\{k\}}  \delta_{N, W_t[k]} \; B[k] \; \prod_{x} \delta(\vec{\nabla} \cdot \vec{k}_{x}) \; .
\label{worldlineZN}
\end{equation}
The Kronecker delta fixes the winding number $W_t[k]$ to the particle number $N$. One can re-construct the grand canonical
partition function with the fugacity series,
$Z_{gc} = \sum_{N}  Z_N \; e^{\, \mu \beta N}$.
The canonical partition sums $Z_N$ describe the system at fixed net particle number $N$ and inverse temperature $\beta = N_t$. 

It is straightforward to modify the Monte Carlo simulation strategy of the grand canonical worldline formulation discussed in the previous
section to the canonical case at fixed particle number $N$. One starts the simulation with an admissible configuration of the $k_{x,\nu}$
that has the desired temporal winding number $W_t[k] = N$. For example $|N|$ straight winding loops in temporal direction, where all 
$k_{x,2}$ on the loop are equal to sign$(N)$, or a single temporally winding loop with all $k_{x,2}$ on the loop equal to $N$. Subsequently
one updates the system as before, but in the Monte Carlo algorithm suppresses the steps that can change the temporal winding number. 
For the algorithm with local updates one simply omits the steps where temporally winding flux is offered to the system, and for the 
worm strategy one uses hard temporal boundary conditions, i.e., the worm is reflected when it tries to cross the last time-slice to connect 
periodically with the first one. All other steps of the two algorithms, such as the update of the auxiliary dual variables $a_{x,\nu}$, are kept
as described in the previous section. 

Having discussed the canonical worldline formulation and the corresponding simulation strategies, we now address the step of converting 
the canonical results to the grand canonical form, i.e., the determination of the chemical potential from the canonical data. We
stress that this step is important for our exploratory study here, where we want to demonstrate that in a practical implementation of
our canonical worldline approach the results agree with the grand canonical picture. In an actual application of the approach
one can of course
equally well stay in the canonical picture and consider the observables as a function of the particle number density $n$.

The chemical potential $\mu(n)$ that corresponds to a given particle density $n =  N/N_s$ is defined as the derivative of the 
free energy density $f_{\lambda}(n)$ with respect to the particle number density $n$:
\begin{equation}
\mu (n) \; = \; \frac{\partial f(n)}{\partial n} \; = \;  \frac{N_s}{2} 
\Bigg[ f \left(\frac{N+1}{N_s}\right) - f \left(\frac{N-1}{N_s}\right)\Bigg] + \mathcal{O}\left(\frac{1}{N^2_s}\right) \; .
\label{mudef}
\end{equation}
In the second step we have discretized the derivative, using $f^\prime(n) = [f(n+\delta n) - f(n-\delta n)]/2\delta n + 
\mathcal{O}( \delta n ^2 )$ with $n = N/N_s$ and $\delta n = 1/N_s$. The discretization effects are suppressed with the square 
of the spatial volume. Thus, in order to obtain $\mu(n)$, we need to compute the free energy density $f$ in the given spatial volume
$N_s$ at particle numbers $N+1$ and $N-1$. 

The free energy density $f(n)$ cannot be calculated directly and usually is determined by integrating a suitable observable 
over some coupling. Here we use the quartic field expectation value $\left\langle |\phi|^4\right \rangle$, 
which in Eq.~(\ref{observables}) is related to the free energy density $f$ via the differential equation 
$\partial f / \partial \lambda = \left\langle |\phi|^4\right \rangle$. Integrating this differential equation over the 
coupling $\lambda$ we obtain
\begin{equation}
f(n) \;\;  = \;\; f(n)\Big|_{\lambda=0} \; + \; \int_{0}^{\lambda} \! \!\! d \lambda^{\prime}  \; \Big\langle |\phi|^4\Big\rangle 
\Big|_{\lambda^{\prime}, N, N_s} \; .
\label{fintegral}
\end{equation}
In the integrand the quartic field expectation value $\langle |\phi|^4\rangle \big |_{\lambda^{\prime}, N, N_s}$ is computed 
in a canonical simulation with particle number $N$, at spatial volume $N_s$ and at different values 
$\lambda^\prime$ for the integration. The upper limit
of the integration is the target coupling $\lambda$ where one wants to evaluate the observables. 
The integration constant for the solution of the differential equation is the free energy density $f(n)\big|_{\lambda=0}$ for
the free case ($\lambda = 0$), again at the given density $n = N/N_s$. At $\lambda = 0$ the free energy density can be 
computed with the help of Fourier transformation and a subsequent projection to the needed particle number $N$. 
We provide a sketch of that calculation in the appendix.
\begin{figure}[t!]
\begin{center}
\hspace*{-12mm}
\includegraphics[height=6.5cm]{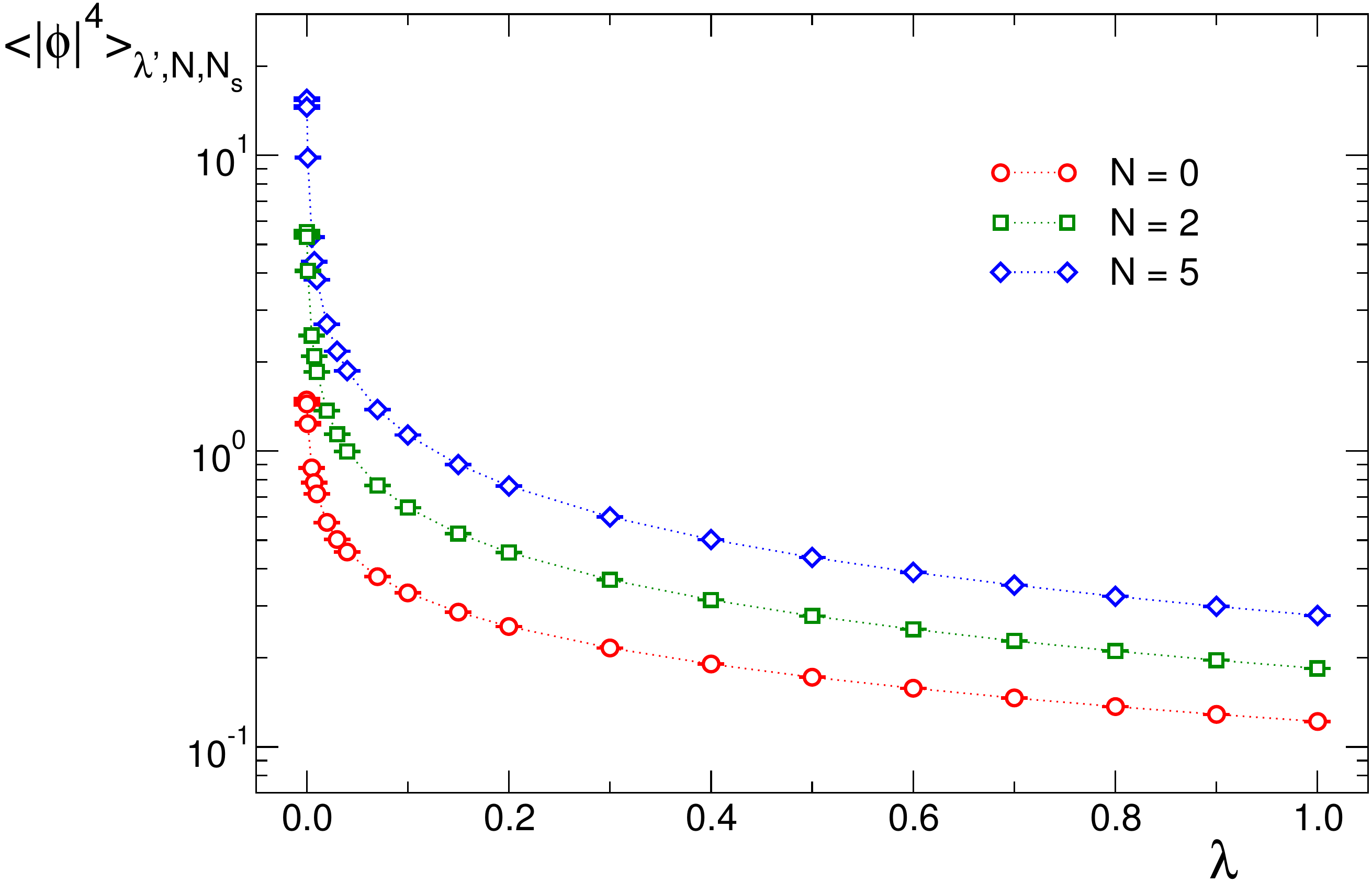}
\end{center}
\vspace*{-6mm}
\caption{The integrand $\langle |\phi|^4\rangle \big |_{\lambda^{\prime}, N, N_s}$ of the free energy density integral 
(\ref{fintegral}) as a function of $\lambda^\prime$ in the interval [0,1]. The data are for $N_s = 10, N_t = 100,
m = 0.1$ and particle numbers $N = 0, 2$ and 5. The numerical results are represented by the symbols connected 
with dotted lines. At $\lambda^\prime = 0$ the observable 
$\langle |\phi|^4\rangle$ is finite -- its value can be computed with Fourier transformation.}
\label{integrand}
\vskip8mm
\begin{center}
\hspace*{-10.5mm}
\includegraphics[height=6.67cm]{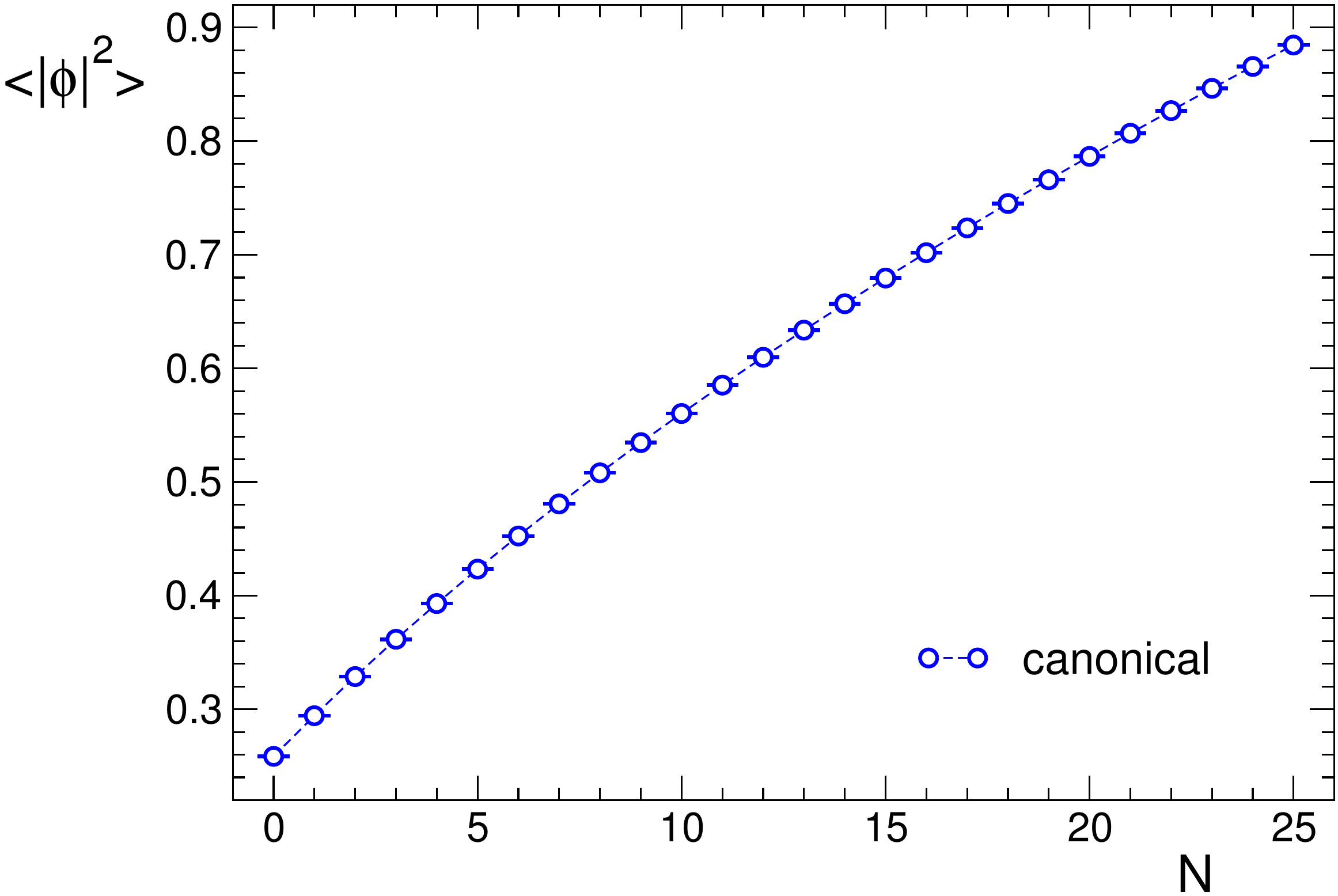}
\end{center}
\vspace*{-6mm}
\caption{Example of a result from a canonical simulation: We show the field expectation value $\langle |\phi|^2\rangle$
as a function of the net particle number $N$ (using $N_s = 10, N_t = 100, m = 0.1$ and $\lambda = 1.0$).}
\label{example}
\end{figure}

The $\lambda^\prime$-integration in (\ref{fintegral}) has to be done numerically and we evaluate
$\langle |\phi|^4\rangle \big |_{\lambda^{\prime}, N, N_s}$ at several values 
$\lambda^\prime \in [0,\lambda]$.  For the tests in this paper we work with $\lambda = 1.0$ and thus
$\lambda^\prime \in [0,1]$. In Fig.~\ref{integrand} we show $\langle |\phi|^4\rangle \big |_{\lambda^{\prime}, N, N_s}$
versus $\lambda^\prime$ for $N_s = 10, N_t = 100, m = 0.1$, and particle numbers $N = 0, 2$ and 5. The values at 
$\lambda^\prime = 0$ are finite and can be computed with Fourier transformation.
The integrands decrease quickly and we found that the numerical integration is very stable. More specifically, 
the integration in (\ref{fintegral}) was done with Mathematica using spline interpolation of the Monte Carlo data.

\begin{figure}[t]
\begin{center}
\includegraphics[height=7.3cm]{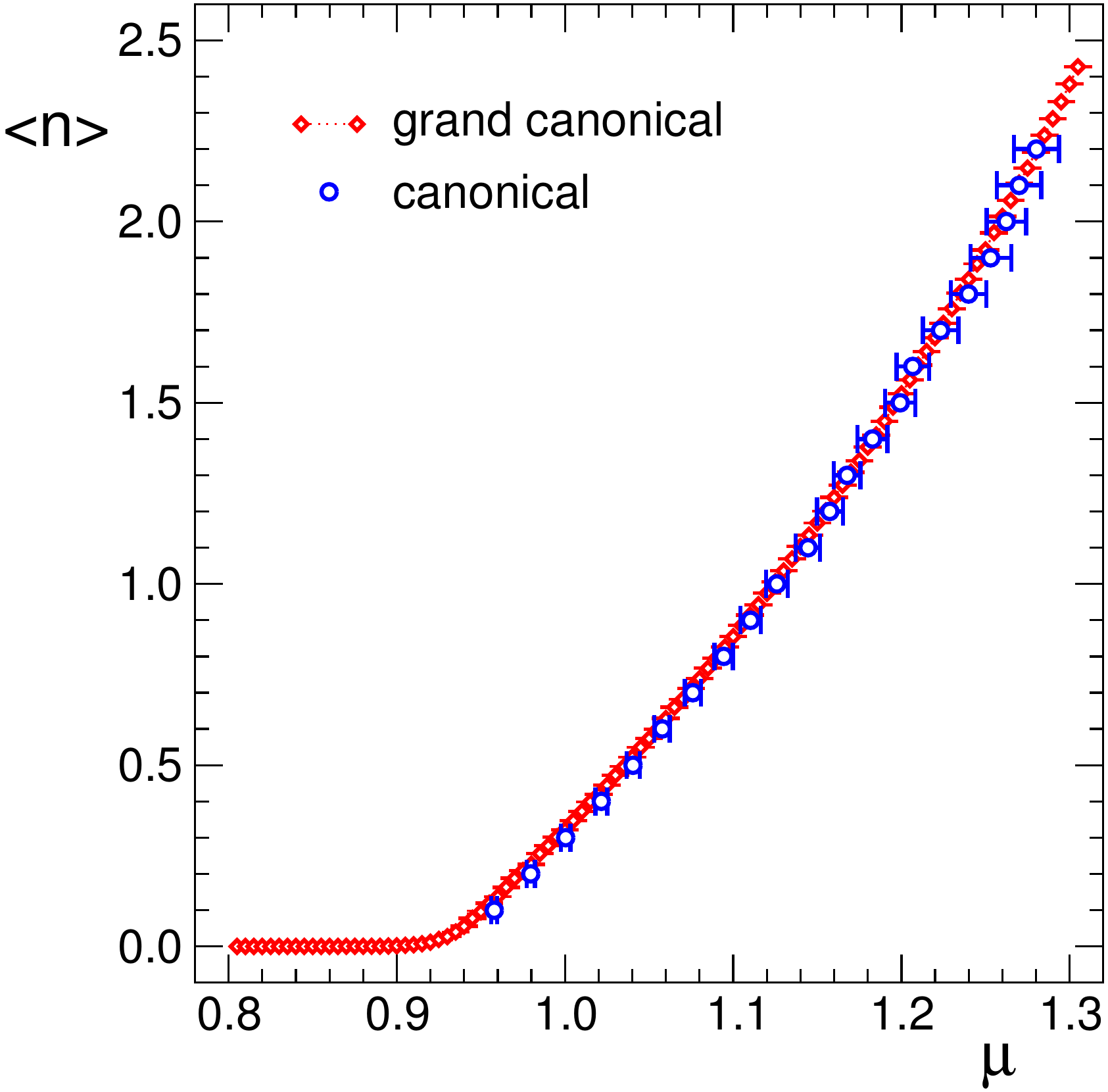}
\hspace{4mm}
\includegraphics[height=7.3cm]{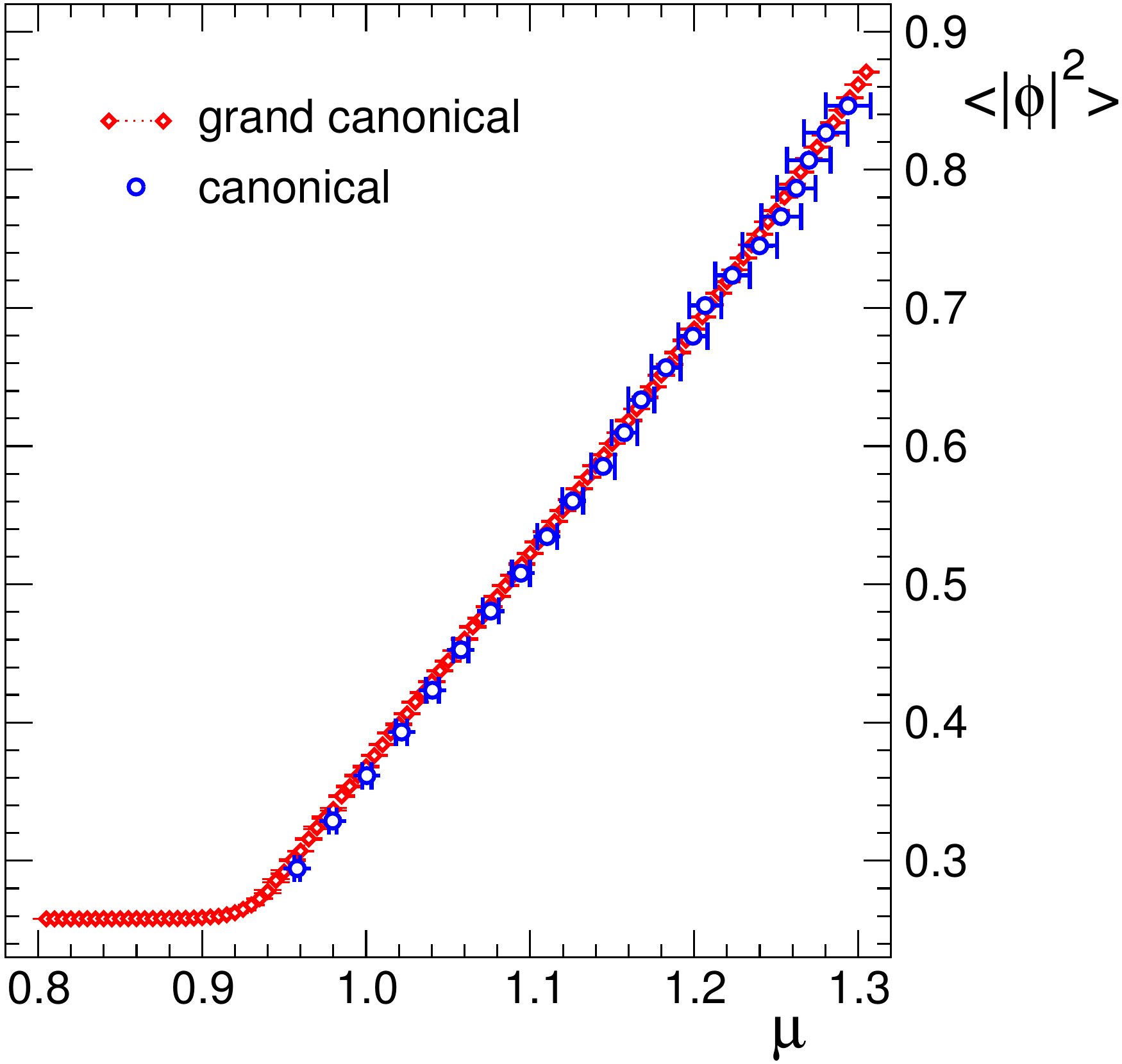}
\end{center}
\vspace*{-6mm}
\caption{Lhs.~plot: the particle number density $\langle n \rangle$
versus $\mu$. We compare the results from the canonical 
approach (blue circles) to those from a grand canonical simulation (red diamonds). 
Rhs.\ plot: the field expectation value $\langle | \phi^2 | \rangle$ as a function of $\mu$ and again we compare 
canonical (blue circles) and grand canonical results (red diamonds).  The data are for $\lambda = 1.0, m = 0.1$,
$N_s = 10, N_t = 100$. In the canonical simulation the chemical potential is an observable such that 
the error bars are horizontal. The canonical $\langle | \phi^2 | \rangle$ results (and all grand canonical results) 
also have vertical error bars which, however, are smaller than the symbols.}
\label{comparison}
\end{figure}

\section{Results and comparison to grand canonical reference data}

Let us now come to the presentation of the results of the canonical simulation and their comparison to the data obtained in a 
grand canonical simulation. For our simulations we compared both algorithmic strategies we discussed, the local updates as well as
the worm algorithm and found excellent agreement of the results. In addition the simulations were cross-checked with the analytical 
calculation available for $\lambda = 0$. The canonical results we show in this section were generated with the local strategy, using 
statistics of $10^5$ configurations separated by $10$ combined sweeps for decorrelation and we used $5 \times 10^4$ sweeps for 
equilibration. The grand canonical reference data were generated with the worm algorithm with $4 \times 10^5$ configurations 
separated by 10 worms for decorrelation and $2 \times 10^5$ worms for equilibration. We worked on different 
lattices with $N_s$ ranging from 8 to 20 and $N_t$ from $8$ to $400$. For the examples shown in this section we used 
$N_s = 10$ and  $N_t = 100$ at $\lambda = 1.0$ and a bare mass parameter of $m = 0.1$. For this setting the physical mass is 
$m_{phys} \sim 0.94$ in lattice units, such that we have a ratio of $T / m_{phys} \sim 0.0094$. At this low temperature $T$ 
we can observe condensation as a function of $\mu$, and thus our choice of parameters tests the canonical
worldline approach in a physically interesting regime. 

We begin the presentation of the numerical data with the canonical results for the field expectation value $\langle | \phi^2 | \rangle$.
In Fig.~\ref{example} we show $\langle | \phi^2 | \rangle$ as a function of the net particle number $N$. This observable is computed 
with the worldline representation (\ref{observables}) for $\langle | \phi^2 | \rangle$, evaluated in the canonical simulations at 
fixed $N$. The results show the expected increase of the field expectation value with increasing $N$. 

A key observable in the canonical approach is of course the chemical potential $\mu(n)$ as a function of the particle number density,
since this quantity is needed for converting the canonical results to the grand canonical form which we use as a cross-check for
our approach. The results from the canonical determination based on (\ref{mudef}) and (\ref{fintegral}) are shown in the lhs.\ plot
of Fig.~\ref{comparison} using blue circles. Note that for the canonical results $\mu$ is the observable which in the figure is plotted
on the horizontal axis and thus has horizontal error bars. The results for $\langle n \rangle$ versus $\mu$ can also be computed 
directly in the grand canonical approach and the corresponding data are shown as red diamonds. The results from the two approaches
agree very well with a slight systematic deviation at small $\langle n \rangle$ which we attribute to the fact that at low particle numbers
the discretization effects of the derivative of the free energy with respect to the particle number are larger. Overall we find that the 
canonical results for $\langle n \rangle$ versus $\mu$ very reliably reproduce the condensation that sets in at $\mu \sim m_{phys}$, 
i.e., when the chemical potential hits the physical mass (below $m_{phys}$ we have $\langle n \rangle = 0$ for vanishing temperature). 

The relation between $\mu$ and $\langle n \rangle$ can now also be used for converting other observables from a canonical 
determination into the grand canonical form. In the rhs.\ plot of Fig.~\ref{comparison} we show the results for $\langle | \phi^2 | \rangle$
versus $\mu$ from the canonical determination as an example (blue circles). 
The canonical data from Fig.~\ref{example} were converted to the grand canonical form
by setting $N = n \,N_s$ and replacing $n$ by $\mu$ using the $n$-$\mu$ relation from the lhs.\ plot of Fig.~\ref{comparison}. 
We compare the results to the direct grand canonical determination (red diamonds) and again find very good 
agreement (with the same small systematic deviation at low densities). Thus also for $\langle | \phi^2 | \rangle$ we can reliably 
describe the condensation phenomenon.

\section{Summary and discussion}

In this letter we have presented an exploratory study for canonical simulations based on worldline representations.  We develop
the approach using the simple example of $\phi^{4}_{2}$ lattice field theory, but it is straightforward to generalize the approach to 
other lattice field theories with a worldline formulation. The key steps are the fact that the net-particle number is given
by the temporal winding number of the worldlines and the identification of suitable Monte Carlo updates that stay within
a fixed sector of winding number. We discuss how the canonical results can be converted into the
grand canonical form via a determination of the $n$-$\mu$ relation from a discretized $n$-derivative of the free 
energy density. 

We numerically implement our canonical approach in $\phi^{4}_{2}$ lattice field theory and systematically compare canonical 
and grand canonical results. We find good agreement of the two approaches in the physically interesting region where 
condensation sets in, i.e., at low temperatures and chemical potential values around the physical mass. This demonstrates 
that the canonical  approach with worldlines can be reliably applied to studying condensation phenomena.

The canonical worldline approach might be particularly useful for fermionic theories with worldlines.
For fermions in a worldline representation obtained from Grassmann integration, the Pauli principle requires that each site of 
the lattice is occupied exactly once with a fermionic element, i.e., a monomer, a dimer or a fermion loop (see, e.g., \cite{schwinger}).
When increasing the chemical potential temporally winding loops start to dominate. A large density of winding loops becomes 
topologically stabilized and the simulation develops long autocorrelations because the Monte Carlo algorithm has to change 
the winding number for properly mapping observables in the condensation region \cite{schwinger2}. 
Here the canonical worldline approach could be a powerful alternative since one can simulate in a fixed winding sector. 
Exploring the new canonical worldline approach for fermions is planned for future studies.

\vskip5mm
\noindent
{\bf Acknowledgements:} We thank Mario Giuliani for many useful discussions. This work is partly supported by
the Austrian Science Fund FWF, grant number I 2886-N27, as well as DFG TR55, ''$\!$Hadron Properties from Lattice QCD''.
Oliver Orasch acknowledges support from the Paul Urban Foundation.

\section*{Appendix}
In this appendix we briefly sketch how one can compute the free energy density $f(n)\big|_{\lambda=0}$ for the
free case ($\lambda = 0$) at a given $N_s$ and $N_t$ and project it to the desired net particle number $N$ such that it 
corresponds to the integration constant in Eq.~(\ref{fintegral}). 

For $\lambda = 0$ the action has only quadratic terms and can be written in the form $S = \sum_{x,y \in \Lambda}
\phi_x^* M_{xy} \phi_y$, with the matrix $M$ given by $M_{xy} = (4+m^2) \delta_{x,y} -
\sum_\nu ( e^{\mu \delta_{\nu,2}} \delta_{x+\hat{\nu},y} + e^{- \mu \delta_{\nu,2}} \delta_{x-\hat{\nu},y})$. 
Thus the grand canonical partition sum (\ref{latticeZ}) is a Gaussian integral with solution $Z_{gc} = (2\pi)^V\! /\! \det M$. The 
Matrix $M$ can be diagonalized with Fourier transformation and the determinant is the product of all Fourier modes.

The next step is to project the determinant to a fixed particle number $N$. This can be done by evaluating the determinant for an 
imaginary chemical potential $\mu = i \varphi / N_t$ and Fourier transformation with respect to $\varphi$, such that we obtain the canonical
partition sum for the free case as $Z_N = (2\pi)^{-1} \int_{-\pi}^\pi d \varphi \, e^{-i \varphi N} \,  Z_{gc} \big|_{\mu = i \varphi / N_t} $.
Evaluating the Fourier modes of $M$ for the choice $\mu = i \varphi / N_t$ and putting things together we find the following expression for 
the canonical partition sum $Z_N$ in the free case (we reorganized the product over the Fourier modes and dropped all overall factors):
\begin{equation}
Z_N \; = \;  \int_{-\pi}^\pi \!\!\!\! d \varphi \;
e^{-i \varphi N} \! \left[ \prod_{k_1 = -N_s/2}^{N_s/2-1} \! \Big( (\eta - 2 c_1)^2 - 4 c_\varphi^2 \Big)
\prod_{k_2=1}^{N_t/2-1} \! \Big[ (\eta - 2 c_1 -2c_\varphi*c_2)^2 - 4 s_\varphi^2 s_2^2 \Big] \right]^{-1} \!\!\!\!\! .
\label{ZNfree}
\end{equation}
We use $\eta = 4 + m^2, c_\varphi = \cos(\varphi/N_t), s_\varphi = \sin(\varphi/N_t), 
c_1 = \cos(2\pi k_1/N_s), c_2 = \cos(2\pi k_2/N_s)$ and $s_2 = \sin(2\pi k_2/N_s)$.  
The expression for $Z_N$ can be evaluated with Mathematica 
and we obtain the free energy density needed for the integration constant in 
Eq.~(\ref{fintegral}) as $f(n)\big|_{\lambda=0} = - \ln Z_N / V$.


\begin{thebibliography}{12}

\bibitem{canonicalQCD}
K.F.~Liu,
  {\sl Finite density algorithm in lattice QCD: A canonical ensemble approach},
  Int.\ J.\ Mod.\ Phys.\ B {\bf 16} (2002) 2017
  [hep-lat/0202026].
%
A.~Alexandru, M.~Faber, I.~Horvath, K.F.~Liu,
  {\sl Lattice QCD at finite density via a new canonical approach},
  Phys.\ Rev.\ D {\bf 72} (2005) 114513
  [hep-lat/0507020].
%
S.~Kratochvila, P.~de Forcrand,
  {\sl The canonical approach to finite density QCD},
  PoS LAT {\bf 2005} (2006) 167
  [hep-lat/0509143].
%
P.~de Forcrand, S.~Kratochvila,
  {\sl Finite density QCD with a canonical approach},
  Nucl.\ Phys.\ Proc.\ Suppl.\  {\bf 153} (2006) 62
  [hep-lat/0602024].
%
  S.~Ejiri,
  {\sl Canonical partition function and finite density phase transition in lattice QCD},
  Phys.\ Rev.\ D {\bf 78} (2008) 074507
  [arXiv:0804.3227].
%
  J.~Danzer, C.~Gattringer,
  {\sl Winding expansion techniques for lattice QCD with chemical potential},
  Phys.\ Rev.\ D {\bf 78} (2008) 114506
  [arXiv:0809.2736].
%
A.~Li, A.~Alexandru, K.F.~Liu, X.~Meng,
  {\sl Finite density phase transition of QCD with $N_f=4$ and $N_f=2$ using canonical ensemble method},
  Phys.\ Rev.\ D {\bf 82} (2010) 054502
  [arXiv:1005.4158].
%
K.~Nagata, A.~Nakamura,
  {\sl Wilson fermion determinant in lattice QCD},
  Phys.\ Rev.\ D {\bf 82} (2010) 094027
  [arXiv:1009.2149].
%
A.~Alexandru, U.~Wenger,
  {\sl QCD at non-zero density and canonical partition functions with Wilson fermions},
  Phys.\ Rev.\ D {\bf 83} (2011) 034502
  [arXiv:1009.2197].
 %
 A.~Li, A.~Alexandru, K.F.~Liu,
  {\sl Critical point of $N_f = 3$ QCD from lattice simulations in the canonical ensemble},
  Phys.\ Rev.\ D {\bf 84} (2011) 071503
  [arXiv:1103.3045].
%
J.~Danzer, C.~Gattringer,
  {\sl Properties of canonical determinants and a test of fugacity expansion for finite density lattice QCD with Wilson fermions},
  Phys.\ Rev.\ D {\bf 86} (2012) 014502
  [arXiv:1204.1020].
%
  C.~Gattringer, H.P.~Schadler,
  {\sl Generalized quark number susceptibilities from fugacity expansion at finite chemical potential for $N_f$ = 2 Wilson fermions},
  Phys.\ Rev.\ D {\bf 91} (2015),  074511
  [arXiv:1411.5133].
%
  A.~Nakamura, S.~Oka, Y.~Taniguchi,
  {\sl QCD phase transition at real chemical potential with canonical approach},
  JHEP {\bf 1602} (2016) 054
  [arXiv:1504.04471].
  
\bibitem{reviews}
  S.~Chandrasekharan,
  {\sl A new computational approach to lattice quantum field theories},
  PoS LATTICE {\bf 2008} (2008) 003
  [arXiv:0810.2419].
%
  P.~de Forcrand,
  {\sl Simulating QCD at finite density},
  PoS LAT {\bf 2009} (2009) 010
  [arXiv:1005.0539].
%
 U.~Wolff,
  {\sl Strong coupling expansion Monte Carlo},
  PoS LATTICE {\bf 2010} (2010) 020
  [arXiv:1009.0657].
%
C.~Gattringer,
{\sl New developments for dual methods in lattice field theory at non-zero density},
PoS LATTICE {\bf 2013} (2013) 002
[arXiv:1401.7788].

\bibitem{phi4}
  C.~Gattringer, T.~Kloiber,
  {\sl Lattice study of the Silver Blaze phenomenon for a charged scalar $\phi^4$ field},
  Nucl.\ Phys.\ B {\bf 869} (2013) 56
  [arXiv:1206.2954].
%
  C.~Gattringer, T.~Kloiber,
  {\sl Spectroscopy in finite density lattice field theory: An exploratory study in the relativistic Bose gas},
  Phys.\ Lett.\ B {\bf 720} (2013) 210
  [arXiv:1212.3770].
%
\bibitem{phi4worm}
  M.~Giuliani, C.~Gattringer,
  {\sl Remarks on the construction of worm algorithms for lattice field theories in worldline representation},
  arXiv:1702.04771.
%
  \bibitem{worm}
  N.~Prokofev, B.~Svistunov,
  {\sl Worm algorithms for classical statistical models},
  Phys.\ Rev.\ Lett.\  {\bf 87} (2001) 160601.

\bibitem{schwinger}
C.~Gattringer, T.~Kloiber, V.~Sazonov,
  {\sl Solving the sign problems of the massless lattice Schwinger model with a dual formulation},
  Nucl.\ Phys.\ B {\bf 897} (2015) 732
  [arXiv:1502.05479]. 
%
C.~Gattringer, T.~Kloiber, V.~K.~Sazonov,
  {\sl Dual representation for massless fermions with chemical potential and U(1) gauge fields},
  PoS LATTICE {\bf 2015} (2016) 195.

\bibitem{schwinger2}
D.~G\"oschl, C.~Gattringer, A.~Lehmann, C.~Weis,
  {\sl Simulation strategies for the massless lattice Schwinger model in the dual formulation},
  arXiv:1708.00649.

\end{thebibliography}
\end{document}